# Mathematical Modeling and Analysis of ZigBee Node Battery Characteristics and Operation

Abla Hussein [1*] and Ghassan Samara [2]

[1]Department of Computer Science, Zarqa University, Jordan
[2]Department of Computer Science, Zarqa University, Jordan

**Abstract**

ZigBee network technology has been used widely in different commercial, medical and industrial applications, and the importance of keeping the network operating at a longer time was the main objective of ZigBee manufacturers. In this paper, ZigBee battery characteristics and operation has been researched extensively and a mathematical modeling has been applied on existed practical data provided by Freescale semiconductors Inc [1], and Farnell [2]. As a result a mathematical optimized formula has been established which describes battery characteristic and voltage behavior as a function of time, and since ZigBee node batteries has been the core objective of this research work, a decline in battery voltage below 50% of battery capacity could influence and degrade ZigBee network performance.

**Key words**: Battery lifetime, ZigBee network, Power consumption, mathematical modeling, Battery model.

## 1. Introduction

ZigBee specifications have been developed by the ZigBee Alliance and it is used mainly for control and sensor networks. Its characteristics depend mainly on IEEE 802.15.4 standards. ZigBee applications are expected to provide low power and low cost connectivity for equipment that needs to be operated on batteries. Network Nodes are required to operate from several months to several years for transmitting and receiving low bit rate signals in wireless networks [3].

IEEE 802.15.4 is a wireless standard that depends mainly on physical layer and Media Access Layer (MAC) of ZigBee Applications and technology. It achieves low power consumption by autonomous low-powered devices. Such devices powered by batteries which requires the ability to go to sleep or shut down; many applications required this type of operations.

Battery usage in ZigBee nodes has the following advantages

1. Easy installation and low-cost
2. Used for Flexible location devices
3. Suitable for scalable network
4. Low duty cycle: duty cycle is the propagation of time interval between transmissions.

However, most of battery power consumption of ZigBee devices is during transmission and reception time, so battery usage in ZigBee networks is very effective and longevity of Battery life time could go to many years of operations and since IEEE 802.15.4 standard using extremely low duty cycle. The transmitting will be in a very small fraction of time and when there are no transmissions and reception, the device will revert to low-power sleep mode to minimize power consumption [4].

WSN have been implemented in many different applications, typical applications usually consist of hundreds of small battery-powered sensor nodes which periodically transmit their sensed data to one or several data sinks which are acting as routers on the their way to the coordinator [5].

Most of the energy conservation research work has been studied the development of the different types of routing protocols aiming to extend node battery life , but unfortunately there are no studies focused on battery characteristics or features for ZigBee network .

In our paper, ZigBee battery characteristics and operation has been researched extensively and a mathematical modeling has been established, in order to be used by system designers and hardware developers to estimate battery capacity requirements and study the expectation lifetime for ZigBee-based nodes.

## 2. Related Work

Wireless technology have been advancing very fast in different applications since it is more convenient and cost effective than wired applications, especially in ZigBee networks and WSN applications; where home automation, medical and industrial are great examples of ZigBee networks.

An experiment has been carried out to observe the activity cycle and consumption in battery energy for a ZigBee network node during start-up, packet transmission, due to loss of connection and during sleep-mode on Texas Instruments CC2520 transceiver and the Freescale MC1322x platform [6].

Another global study has been presented on energy consideration based on IEEE 802.15.4 technology, the experiment carried out on node prototype powered by Lithium batteries, the authors concluded that the lifetime announced by hardware manufacturer was not studied well specially when used for networking devices, also the have shown that for variable loads the battery chemistry play big part in reducing node performance life time [7]

Other researchers looked and examined different battery models that describe battery capacity utilization based on battery discharged for different parts of circuits and loads where they could consume battery power. Their study reflects a qualitative insight into how a battery's capacity is influenced by multiple factors [8].

The energy requirements based on application activities has been proposed as a model in more study to extend battery life time, a generic low duty-cycle ZigBee® system used as a case study. Experimental results has been carried out and an estimation of 3% error in average was detected between model values compared to experimentations values from which they concluded that battery- driven system design solution's can address energy challenges [9].

Temperature is another factor that influences the performance of batteries in general, and since batteries are composed of many chemical materials, warmer temperatures will decrease battery life time because heat will speed up chemical reactions that could cause corrosion of the internal electrodes [10].

Jennic technology for a changing world, in their publication have provided guidance when using a coin cell to power a device based on the Jennic JN513x wireless microcontroller. An extensive study has been carried out to study battery lifetime under maximum current consumption conditions for Sanyo CR2032 coin battery [11].

Some other research work has been concentrated on improvement and energy optimization for different ZigBee routing algorithms. Authors in [14] used ZigBee AODV mesh network to send and receive route request, the network has been divided into several logical clusters for the purpose of reducing flooding route requests. Their simulations results show improvements in battery capacity and computing power limitation, the authors concentrate on ZigBee topology but do not concentrate on battery capacity as such.

Authors in [15] have focused on reducing energy consumption by using a fussy logic based metric in AODV ZigBee mesh network which is considered as decision making for the discovery process and they concluded that the simulation results displayed a reduction of 70% in energy consumption by cutting down the number of messages sent; thus reducing collision and overhead.

ZigBee node battery runtime (discharge) or life time in hours could be calculated using Equation (5), Peukert has started the calculation of battery discharge time many years ago, Peukert is German scientist (1897) which he expressed the battery capacity in terms of the rate at which it is discharged, as the rate increases, and the battery's available capacity decreases [12].

### 3. Battery Power Consumption

Battery energy of ZigBee network nodes will be consumed and exhausted as a function of power consumption in the various electronic components and systems of the network nodes which is resulted from transmitting, receiving signals, sleeping and during node idle condition, the following analysis will discuss the major components that contribute to energy consumption.

The following formula represents the battery total power consumption of one node of ZigBee network:

$$P_{con} = P_{tx} + P_{rx} + P_{Sleep} + P_{idle} \qquad (1)$$

Where:
$P_{Con}$ = total power consumption.
$P_{tx}$ = power consumption due to transmitted signal
$P_{rx}$ = power consumption due to received signal,
$P_{Sleep}$ = power consumption during sleep state, where the router is in sleep mode, and
$P_{idl}$ = power consumption due to Idle state, where there is no packets are transmitted or received.

Where the relationship between energy and power is represented by:

$$P = E/t \qquad (2)$$

Where $P$ is power measured in Watts, $E$ is energy measured in Joules, and $t$ is time measured in hours, so the $E_{con}$ is:

$$E_{con} = P_{con} t \qquad (3)$$

Where $E_{con}$ = Consumed Energy
After t time, the residual energy $E_{res}$ (energy left in battery) is represented by:

$$E_{res} = E_{int} - E_{con}, \qquad (4)$$

Where $E_{int}$ is the initial energy of the battery, in industry, battery strength is not measured in Joules, it is measure in volts and its capacity measured in mAH.

### 4. Battery Life Time Calculation

The battery lifetime in hours could be calculated using the following formula:

$$T = I_C/I^n \qquad (5)$$

Where:
T = Battery life time in Hours

$I_C$ = Battery Capacity in mAH, I = Load current in mA

n= Peukert's exponent, it ranges from 1 to 1.3, where 1 is the nominal value.

Peukert exponent varies and increases in its value according to the age and type of the battery, in our study we concentrate on coin batteries only. We will use Equation (5) to calculate the battery life time at n=1 for Freescale practical data [1].

**Table 1: Characteristic Data for Battery Life Time Calculation**

| Characteristic | value |
|---|---|
| Peukert's exponent "n" | 1 |
| The Battery capacity | 220mAH, |
| The standard load current which is used for transmission and reception | 0.2 mA |
| The load current that is used due to sleeping and idle states | 0.048 mA |
| The total current | 0.248 mA |
| Theoretical Battery Life (in Hours) | 220 mAH / 0.248 mA= 887 hours |
| The number of hours that is provided by Freescale data sheet | 848 hours |

## 5. Battery models

There are many studies regarding battery models, authors in [10], summarizes battery models, starting with the 1st generation model which is described as a linear storage of current. It is simple model which analyzes battery capacity, but it is not sufficient to characterize battery behavior in real life. 2nd generation model is where batteries operate under constant load condition, where the battery's capacity is reduced as the discharge rate increases from which factor "k" is introduced and it is defined as "the ratio of battery effective capacity divided by battery maximum capacity". This model helps to explain how WSN battery will function under discharge rate changing due to transmitting and receiving signals, so an accurate lifetime characteristics could be predicted. Peukert's law battery model is used to predict battery lifetime in which it is similar to the 2nd generation where nonlinear characteristics have been considered for battery operation. According to Peukert's law which is given by Equation (5), it is simple to understand the battery lifetime behavior under constant load condition. In the 3rd generation model, the battery will discharge at high rate, and if the battery is cut off during the discharge, diffusion and transport rate of the battery's active materials will catch up with the depletion of the materials, this phenomenon is called relaxation effect, and it gives the battery a chance to recover its lost capacity during discharge.

## 6. Mathematical Modeling of Battery Characteristics

Mathematical modeling is a process of investigating relationship among different variables. Regression techniques [13] are among the tools that could be used to establish a mathematical formula from the knowledge of practical data where voltage is dependent on time.

A mathematical modeled formula has been proposed for battery characteristics in which it represents the voltage decay of the voltage (battery power consumption behavior) as a function of time, and a 4th degree order polynomial has been proposed for this purpose as shown in Equation (6) .

In this research work we used and implemented MatLab curve fitting techniques to predict the formula that fits the practical data, where all coefficients have been calculated.

Freescale and Farnell battery practical data have been used for this purpose at different loads, which shows a good fitting between the practical data and the proposed 4th degree order polynomial. Freescale and Farnell practical data were very informative and detailed, in which the practical data has been measured for 7.5 kohms and 15 kohms.

$$V_{out} = \sum a_m t^m \qquad (6)$$

Where $m = 0 \rightarrow 4$

$V_{out}$ = battery capacity in volts
$t^m$ = time in hours
$a^m$ = coefficient (constant)
m = polynomial power degree (it is 4th in our equation)

Expanding the above formula we get [12]:

$$V_{out} = a_0 + a_1 t^1 + a_2 t^2 + a_3 t^3 + a_4 t^4 \qquad (7)$$

Where $a_0, a_1, a_2, a_3, a_4$ are coefficient that could be evaluated with the aid of practical values of battery operation using any regression or curve fitting technique, in this study MatLab has been used for this purpose. Equation (7) could be used for fitting purposes using the practical data of voltage decline as a function of time using any software that could be used for fitting purposes.

## 7. Modeling Results

Figure (1) shows the curve fitting of Freescale data at 1 second transmission (the transmission in seconds has been converted to hours for better understanding of the model and more convenient in plotting and calculation). In curve fitting technique, many formulas could be used for better fitting and that depends on researches and their background in selecting the suitable formula, in our study 4th order polynomial has been selected for better fitting since the practical data variation comes near to this specific polynomial.

As shown in Figure (1), the Freescale practical date (shown as +) and the solid line which represent the 4th polynomial formula, as seen the fitting is excellent and the decline in voltage value appears deeply and sharp at 850 hours of operation at which the voltage could go below 50% of its strength.

Equation (8) shows the polynomial formula that fits Freescale data at 1s.

$V_{out} = 3.16 + 0.00309 t^1 + 1.125E-5 t^2 – 1.36E-8 t^3 + 4.255 E-12 t^4 \qquad (8)$

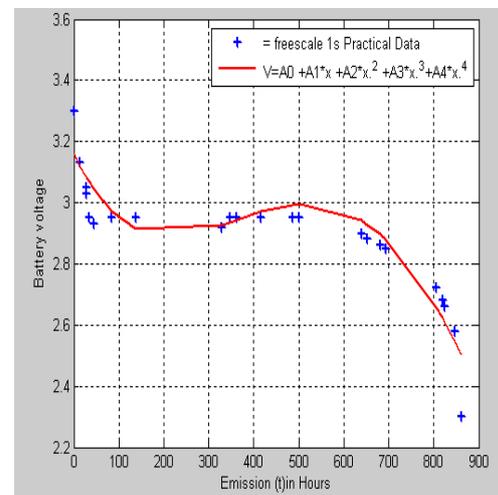

Figure 1: Curve fitting of Freescale practical data.

Figure (2) represents the curve fitting of Farnell practical data at 15kohms (circle points) with the polynomial equation represented in Equation (9). The second curve represent the curve fitting of Farnell practical data at 7.5kohms (triangle points) with the polynomial equation (10) , As seen in Figure (2), it is noticed that at 7.5 kohms the decline in voltage values is faster; almost half the time required for the decline of the 15 kohms results.

From figure (2) we conclude that battery voltage will decline faster for smaller load values.

$V_{out} = 3.292 - 0.0012 t^1 - 2.464E-6 t^2 + 8.92E-9 t^3 - 6.3E-12 t^4 \qquad (9)$

$V_{out} = 3.292 - 0.0015 t^1 + 1.32E-5 t^2 + 4.63E-9 t^3 – 4.17E-11 t^4 \qquad (10)$

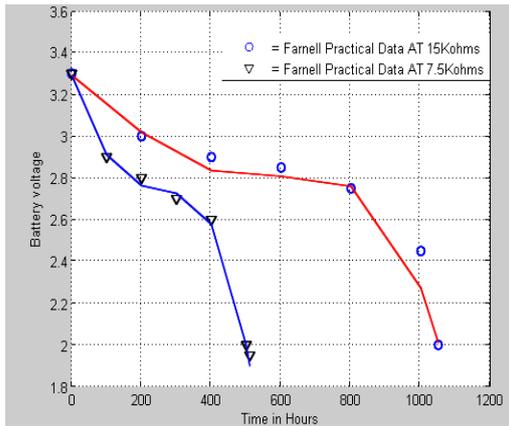

Figure 2. Curve fitting of Farnell practical data at 15k ohms and 7.5 k ohms.

## 8. Power Consumption Relationship

The power relationship between voltage and load is given by:

$$P = V_{out}^2/R \qquad (11)$$

Where:
P = power in Watts,
$V_{out}$ = Battery voltage in Volts,
R = Load in kilo ohms

Formula (11) represents a relationship between R and *P*, as *R* decreases, the power will increase, which means that more current will be drawn by the load, accordingly, and the battery will start discharging faster; as such the battery lifetime will be getting shorter, and vice versa. As the load resistance increase, less power will be consumed and a longer life will be resulted.

If there is no load connected to the battery that does not mean the battery lifetime is unlimited. There is a percentage of battery energy consumption which is less that 1% per year will be reduced as mentioned in most battery manufacturer product data sheets.

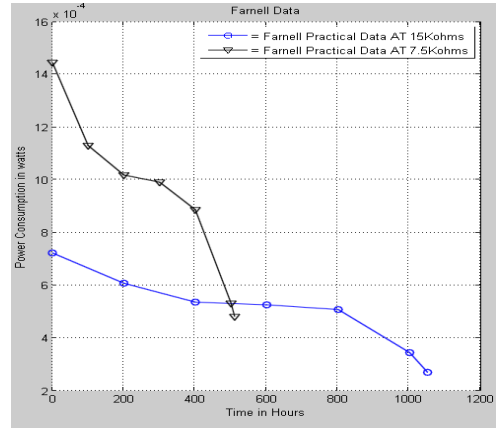

Figure (3) Battery power consumption for Farnell data at 15k ohms and 7.5k ohms

Figure (3) represents a curve of power consumption in Watts versus time in hours for two different loads. At load 7.5k ohms it is noticed that the battery will discharge faster and the battery voltage value will decline from 3.3 Volts to 2 volts in 500 hours, while at 15k ohms, the battery will discharge slower and its voltage value will be reduced from 3.3 V to 2 V in 1050 hours which is twice of the 7.5k ohms load, this concludes our discussion of power consumption in loads and its effect of reducing battery voltage.

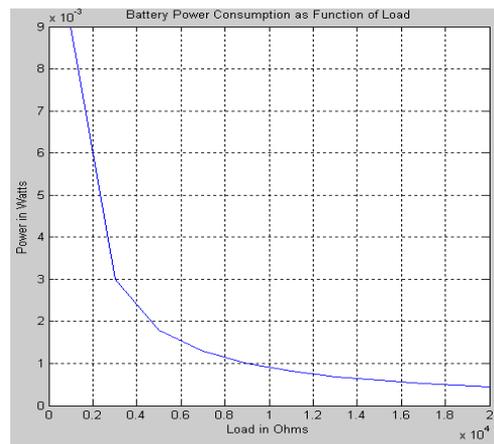

Figure(4) Battery power consumption at different loads

Figure (4) shows the power consumption relationship in watts for different loads varies from 1k ohms to 20K ohms using power equation (11). The concept will apply here precisely, as seen in the figure for a larger load value there will be lesser power consumption, while for small load value, there will be more power consumption. This concept is similar to car battery, if there is a short circuit in the car wiring circuits or car cabinet light was lit all night, the battery capacity will be consumed fast and the voltage will decline below 12 V and the car wouldn't start when you wake up in the morning.

## 9. Conclusion

WSN and ZigBee technology have been implemented widely for many purposes in Industry, hospitals, homes, .etc. and since Batteries are the main power for most of the different application nodes in most of WSN and ZigBee networks, an extensive research work and study are required for the purpose of network longevity.

The technique in this paper is used to predict the relationship of battery voltage decay as a function of time, where voltage is considered as the dependent variable and elapsed time of battery usage is the independent variable. After evaluating the polynomial coefficients that play a big part in the outcomes, the predicted formula could help study and understand battery voltage decline where many factors affects its power discharge and lifetime ending. Some factors like the transmission and receiving of signals, device sleeping mode, idle condition and load values.

The predicted mathematical model is carried out at constant room temperature, and at different load conditions. As a result of such the optimized mathematical model was accurate enough to reflect the minimal difference between the mathematical model values and the practical data; which is considered as a reference to calculate voltage discharge value as a function of time.

Our conclusion of the power relationship and battery discharge time was conclusive to where a smaller load can consume more energy of the battery and send the battery towards a faster discharge situation and a reduction in battery voltage level which is considered very critical to ZigBee nodes operations.

For future work, many factors could be taken into consideration like temperature change and internal battery resistance. Also a precise value of Peukert exponent could be calculated for this kind of coin batteries for a better lifetime calculation.